\documentclass[twocolumn,english,aps,prb,floatfix,showpacs,showkeys,amsmath,amssymb,eqsecnum,superscriptaddress]{revtex4-1}
\usepackage[T3,T1]{fontenc}
\usepackage[latin1]{inputenc}
\usepackage{color}
\usepackage{babel}
\usepackage{amsmath}
\usepackage{amssymb}
\usepackage{graphicx}
\usepackage[unicode=true,
 bookmarks=false,
 breaklinks=false,pdfborder={0 0 1},backref=false,colorlinks=true]
 {hyperref}
\hypersetup{
 pdfauthor={Plekhanov;Evgeny},
 pdftex,linkcolor=blue,citecolor=blue,filecolor=blue,urlcolor=blue,pdftitle=,pdfsubject=,pdfkeywords=}

\makeatletter

\providecommand{\tabularnewline}{\\}

\usepackage[noenc]{tipa}
\usepackage{tipx}
\usepackage{pifont}
\usepackage{eurosym}
\usepackage{wasysym}
\usepackage{amsfonts}\usepackage{textcomp}
\usepackage{array}
\usepackage{supertabular}
\usepackage{hhline}

\makeatother

\begin{document}
\title{Efficient magnetic superstructure optimization with $\Theta\Phi$}
\author{Evgeny A. Plekhanov}
\email{evgeny.plekhanov@kcl.ac.uk}

\affiliation{King's College London, Theory and Simulation of Condensed Matter (TSCM),
The Strand, London WC2R 2LS, UK}
\affiliation{A.N. Frumkin Institute of Physical Chemistry and Electrochemistry
of RAS, Moscow, Russia}
\author{Andrei L. Tchougr{\'e}eff}
\affiliation{A.N. Frumkin Institute of Physical Chemistry and Electrochemistry
of RAS, Moscow, Russia}
\date{\today}
\begin{abstract}
Simulating the incommensurate spin density waves (ISDW) states is
not a simple task within the standard \emph{ab-initio} methods. Moreover,
in the context of new material discovery, there is a need for fast
and reliable tool capable to scan and optimize the total energy as
a function of the pitch vector, thus allowing to automatize the search
for new materials. In this paper we show how the ISDW can be efficiently
obtained within the recently released $\Theta\Phi$ program. We illustrate
this on an example of the single orbital Hubbard model and of $\gamma$-Fe,
where the ISDW emerge within the mean-field approximation and by using
the twisted boundary conditions. We show the excellent agreement of
the $\Theta\Phi$ with the previously published ones and discuss possible
extensions. Finally, we generalize the previously given framework
for spin quantization axis rotation to the most general case of spin-dependent
hopping matrix elements. 
\end{abstract}
\maketitle

\section[Introduction]{Introduction}

Incommensurate spin structures (ISS) appear in various contexts of
condensed matter physics: frustrated spin systems\citep{1Nersesyan1998,2Plekhanov2006},
cuprates high-temperature superconductors\citep{3Thurston1989,4Plekhanov2011,5Avella2011}
and strongly correlated electronic systems in general. In particular,
the question of how the ISS appear as a consequence of strong electron
repulsion, or spin exchange is very interesting, since the ISS come
out as a result of a subtle balance between the kinetic energy loss
and potential energy gain. In Ref.\onlinecite{6Arrigoni1991}, it
was shown how the ISS emerge out of the on-site repulsion, at least
within the mean-field approximation. On the other hand, in Refs.\onlinecite{6Arrigoni1991,7Mryasov1991,8Sandratskii1991,9Sandratskii1991},
it was shown how an arbitrary ISS pitch vector can be treated without
increasing the unit vector size by introducing the twisted boundary
conditions instead of the periodic ones. It was shown therein, that
at least at the mean-field level, ISS emerge as the most stable phase
at the intermediate interaction strength and away from half-filling.

On the other hand, recently, our program \textit{$\Theta\Phi$}\citep{10Plekhanov2017,11Plekhanov2020},
allows for ISS, superconductivity of arbitrary order and Resonating
Valence Bond (RVB) states in multi-orbital electronic systems at finite
temperature. In addition, \textit{$\Theta\Phi$} is capable to import
the hopping parameters from major \textit{ab-initio} codes by means
of \textsc{wannier90}\citep{12Mostofi2008} and \textsc{lobster}\citep{13Deringer2011,14Maintz2016,15Tchougreeff2013}
programs, which makes it possible to perform practically \textit{ab-initio}
strongly correlated magnetic or superconducting calculations.

The scope of this paper is twofold. Firstly, we test and benchmark
the capabilities of \textit{$\Theta\Phi$} applying it to the ISS
in single orbital Hubbard model and in $\gamma$-Fe and compare our
findings with those obtained previously and independently within the
approaches and programs different from \textit{$\Theta\Phi$}. Secondly,
in this paper, we present a general framework for the spin quantization
axis rotation in a general case when no assumptions are made on the
spin dependence of the Hamiltonian's hopping matrix elements. This
general framework extends the one already presented in the Ref.\onlinecite{11Plekhanov2020}
and allows to treat the cases when the hopping is spin-dependent and
contains the spin-flip terms, thus permitting simulations of systems
with spin-orbit coupling and explicit break of time-reversal symmetry.

This paper is organized as follows: we present the methods used in
Sec.2 and Appendix A, the results of our \textit{$\Theta\Phi$} calculations
are shown in Sec.3, while Discussion and Conclusions are given in
Sec.4.

\section[Methods]{Methods}

Recently proposed program $\Theta\Phi$\citep{11Plekhanov2020} allows to fulfill
two of the most important extensions of the essentially single-particle
mean-field paradigm of the computational solid state physics: the
admission of the Bardeen--Cooper--Schrieffer electronic ground state
and allowance of the magnetically ordered states with an arbitrary
superstructure (pitch) wave vector. Both features are implemented
in the context of multi-band systems and permit the interface with
the solid state quantum physics packages eventually providing access
to the first-principles estimates of the relevant matrix elements
of the model Hamiltonians derived from the standard DFT calculations.
In the Ref.\onlinecite{11Plekhanov2020}, we have presented a general
approach to the spin quantization axis rotation, which correctly works
in the case of multi-orbital spin-independent hopping matrix elements.
In the present paper, we extend the spin quantization axis rotation
method to a general case of arbitrary (Hermitian) hopping. The details
of the derivation are given in the Appendix A, while the brief summary
is shown below:
\begin{itemize}
\item At a finite pitch vector $\mathbf{\mathbf{Q}}$, a hoping matrix $t_{\mathit{ij}}(k,k^{\prime})\delta(k-k^{\prime})$
\ is transformed into a new matrix $\widetilde{t}_{\mathit{ij}}(k,k^{\prime})\delta(k-k^{\prime}\pm\mathbf{Q})$
\ in which the states with momentum \textit{k} and spin e.g. ``up''
are connected to the states with the momentum $k+\mathbf{Q}$ \ and
the spin ``down'' and vice-versa.
\item The translational invariance, ``broken'' by imposing the incommensurate
spin spiral with the pitch vector $\mathbf{Q}$, can be restored if
we shift the ``down'' states by $\mathbf{Q}$ prior to the solution
of the secular equation.
\item The interaction terms like Coulomb repulsion and Heisenberg exchange,
which are typically fourth rang tensors, have to be transformed according
to the formulas given in Ref.\onlinecite{11Plekhanov2020}, and this
task is facilitated by the fact that the most important contributions
to them are local \textit{i.e.} do not extend outside of the unit
cell.
\item In the case when the hopping term is spin-independent, the $\mathbf{Q}$-dependence
only comes from the interaction term, and the ISS state is stabilized
if the energy gain from the ``spiralizing'' the interaction is greater
than the energy lost by the kinetic energy.
\item In the case when the hopping term is spin-dependent, the kinetic energy
becomes $\mathbf{Q}$-dependent, although the main contribution to
the stabilization of the ISS is still expected to come from the interaction
terms.
\end{itemize}
The mean-field self-consistency workflow proceeds as usual, with the modified
Hamiltonian, and, at self-consistency, the internal energy $E(\mathbf{Q})$
\ as a function of $\mathbf{Q}$ is obtained.

\section{Results and Benchmarks}

\subsection{Incommensurate spin spirals in Hubbard model}

The single orbital Hubbard model considered in this paper has the
following Hamiltonian:

\begin{equation}
H=-t\sum_{\langle mn\rangle}\left(c_{\mathit{n\sigma}}^{\dagger}c_{\mathit{m\sigma}}^{\phantom{\dagger}}+\text{H.c.}\right)+U\sum_{n}n_{n\uparrow}n_{n\downarrow}-\mu\sum_{\mathit{n\sigma}}n_{\mathit{n\sigma}}
\label{Ham}
\end{equation}
Here $c_{\mathit{n\sigma}}$ \ is the electron annihilation operator
on site $n$ \ with spin $\sigma$, $n_{\mathit{n\sigma}}=c_{\mathit{n\sigma}}^{\dagger}c_{\mathit{n\sigma}}$
\ is the electron occupation operator, $t$ \ is the nearest neighbor
hopping parameter, $U$ \ is the onsite Coulomb repulsion, while
$\mu$ \ is the chemical potential. In the Hamiltonian~(\ref{Ham}), the electrons
move on a square lattice and only nearest neighbor hopping is considered,
which is emphasized by the notation $\left\langle nm\right\rangle $.
Finally, $\mu$ \ is the system's chemical potential, which enforces
the correct number of particles in the system. The calculations are
performed at a finite temperature $T$ \ which was typically kept
low enough ($T=10^{-4}t$) and on a dense 2D \textit{k}-point grid
of $400\times400$ \ points.

The Hamiltonian~(\ref{Ham}) cannot be solved exactly as it is, however, in
the atomic limit ($t=0$) it does allow for the exact solution, revealing
a rich phase diagram with charge and spin orderings, which rapidly
evolve as a function of temperature\citep{16Mancini2012,17Mancini2013,18Mancini2012,19Mancini2013}.

In this work, we consider the incommensurate spin density waves with
the pitch vector defined as:

\begin{equation}
\mathbf{Q}=\left(\pi,\pi\right)+\delta\left(1,1\right).
\end{equation}
This phase is referred to as the $(1,1)$ phase in Ref.\onlinecite{6Arrigoni1991}
and it is assumed that $\delta\ll1$.

Within \textit{$\Theta\Phi$}, the calculations are performed by using
the extended density matrix $\rho$ as the self-consistency procedure
variable, so that the internal energy $E(\mathbf{Q})$ at each pitch
vector $\mathbf{Q}$ is at self-consistency. In \textit{$\Theta\Phi$},
the density matrix is of the most general form, allowing for incommensurate
spin-density waves, superconductivity, Resonating Valence Bonds (RVB).
The extended density matrix is defined through the basic fermionic
operators as follows:

\begin{equation}
\rho_{n,m}\left(\tau\right)=\delta_{\tau,0}\delta_{n,m}-\langle\Psi_{n,R}^{\dagger}\Psi_{m,R+\tau}^{\dagger}\rangle,
\end{equation}

and the operatorial basis $\Psi_{R}^{\dagger}$ \ for a single-orbital
model is defined as:

\begin{equation}
\Psi_{R}^{\dagger}=\left(c_{R\downarrow}^{\dagger},c_{R\uparrow}^{\dagger},c_{R\downarrow}^{\dagger},c_{R\uparrow}^{\dagger}\right).
\end{equation}

In the particular case of a single band Hubbard model with spin-density
waves and without superconductivity and RVB, the structure of $\rho$
is of the following form:

\begin{equation}
\rho=\left(\begin{matrix}\frac{n-\sigma}{2} & 0 & 0 & 0\\
0 & \frac{n+\sigma}{2} & 0 & 0\\
0 & 0 & 1-\frac{n-\sigma}{2} & 0\\
0 & 0 & 0 & 1-\frac{n+\sigma}{2}
\end{matrix}\right).\label{eq:rhoHub}
\end{equation}

Here, $n$ is the average site occupation, while $\sigma$ is the
average site magnetization. One can see from Eq.\eqref{eq:rhoHub}
that this type of $\rho$ describes a site with occupation imbalance
between `up' and `down' channels. This would normally correspond to
the ferromagnetic order, however, thanks to the method of spin quantization
axis rotation, proposed in Ref.\onlinecite{6Arrigoni1991,7Mryasov1991}
and implemented in \textit{$\Theta\Phi$}, an arbitrary spin-density
wave pitch vector can be introduced and treated at the same computational
cost as the simplest ferromagnetic phase.

\begin{figure}
\includegraphics[angle=270,width=1\columnwidth]{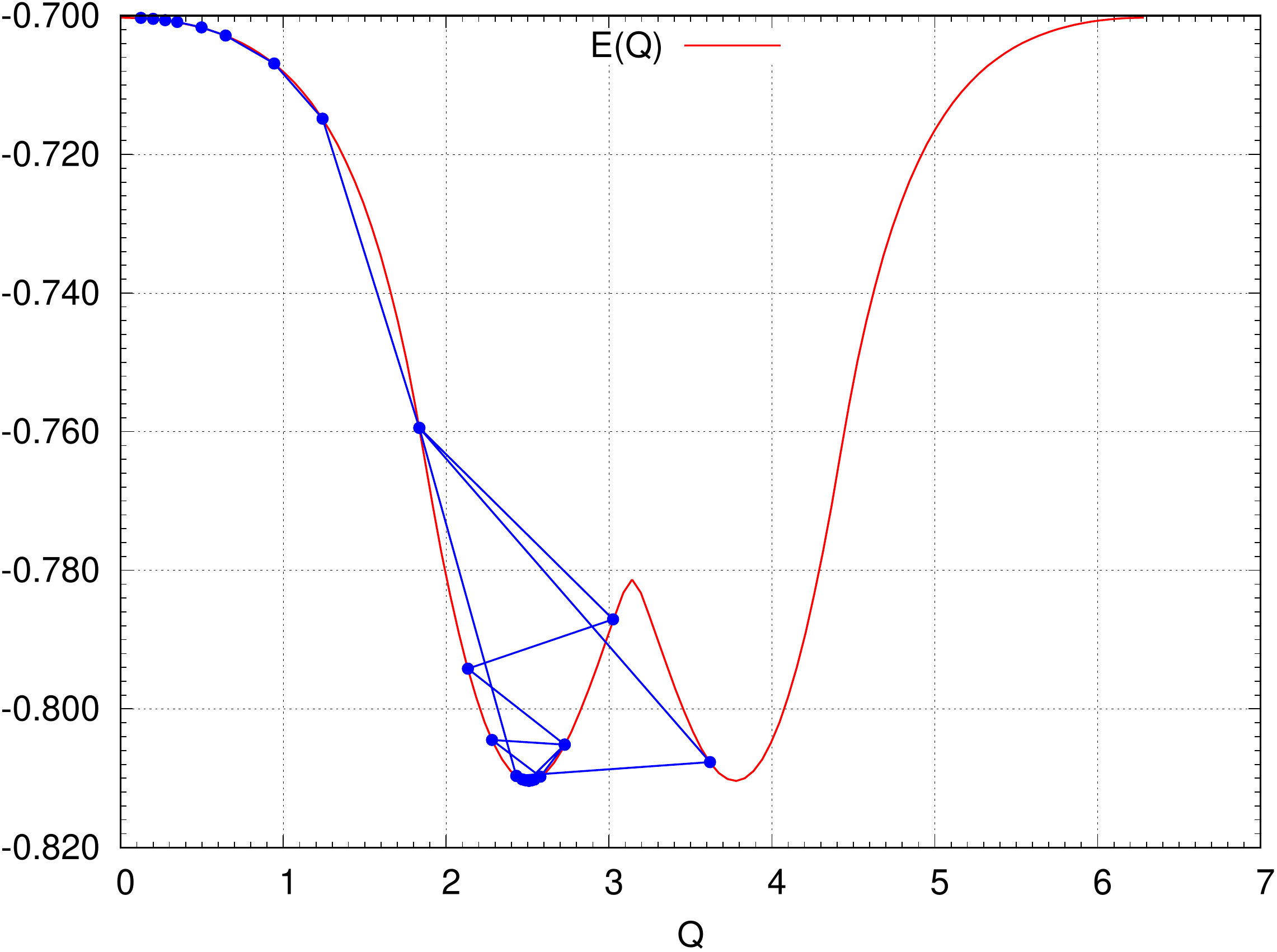}\caption{ \label{fig:1}A typical internal energy profile $E(\mathbf{Q})$\ $U=5t$,
$\delta=0.15$, $T=0.0001t$. $E(\mathbf{Q})$ \ has two local minima
symmetric with respect to $(\pi,\pi)$, which is a local maximum.
The blue line shows the simplex optimization within $\Theta\Phi$.}
\end{figure}

The multidimensional optimization built in \textit{$\Theta\Phi$}
(simplex algorithm) allows to optimize efficiently the internal energy
of the system $E(\mathbf{Q})$ \ as a function of the pitch vector
$\mathbf{Q}$. A typical internal energy profile and the progress
of the minimization is shown in Figure\ref{fig:1}. We notice that
the commensurate antiferromagnetism with the pitch vector $\left(\pi,\pi\right)$
corresponds to a local maximum of $E(\mathbf{Q})$, and the minimum
of $E(\mathbf{Q})$ \ is two-fold degenerate with the minima located
symmetrically with respect to $\left(\pi,\pi\right)$. The optimization
procedure shown in Figure\ref{fig:1} corresponds to a single set
of the Hamiltonian parameters U and $\delta$. We have performed a
scan in this parameter space in order to benchmark our implementation
against the data published in Ref.\onlinecite{6Arrigoni1991}. The
comparison can be seen in Figure\ref{fig:2}. We notice a very good
overall agreement between the two results which validates the use
of \textit{$\Theta\Phi$} in the context of ISS. At largest $t/U$
considered $(t/U=0.2)$, the offset as a function $\delta$ starts
linearly from zero and reaches saturation around $\delta=0.25$. At
a smaller value of $t/U=0.1$ ($U=10t$), the offset monotonically
grows from zero to approximately $a\Delta\mathbf{Q}/\pi=0.8$.

\begin{figure}
   \includegraphics[width=1\columnwidth]{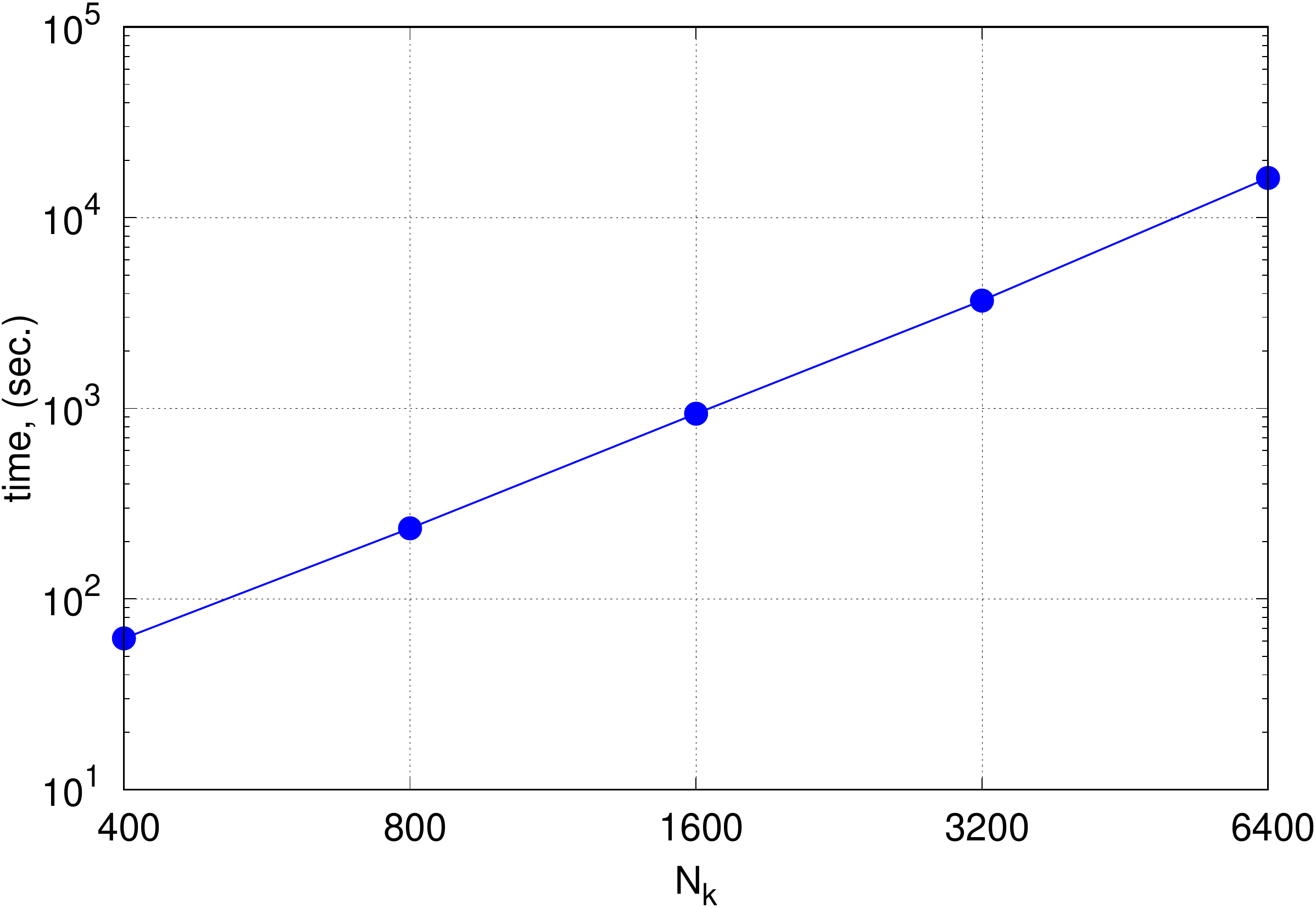} 
   \caption{\label{fig:bench}
	  A benchmark of \textit{$\Theta\Phi$}, showing the dependence
		 of the time for single self-consistency solution on the number of $k$-points along each of the
		 dimensions of a 2D grid in the Brillouin zone. Note the logarithmic scale on both axes.
		 Here, $N_k$ is the number of $k$-points along each of the dimensions of a 2D grid in the
		 Brillouin zone, so that in each grid there are $N_k\times N_k$ points. $U=10t$, $T=0.01t$,
		 ISS phase at $\mathbf{Q}=0$.
   }
\end{figure}

\begin{figure*}
\includegraphics[viewport=0bp 13.08571bp 867bp 687bp,clip,width=0.48\textwidth]{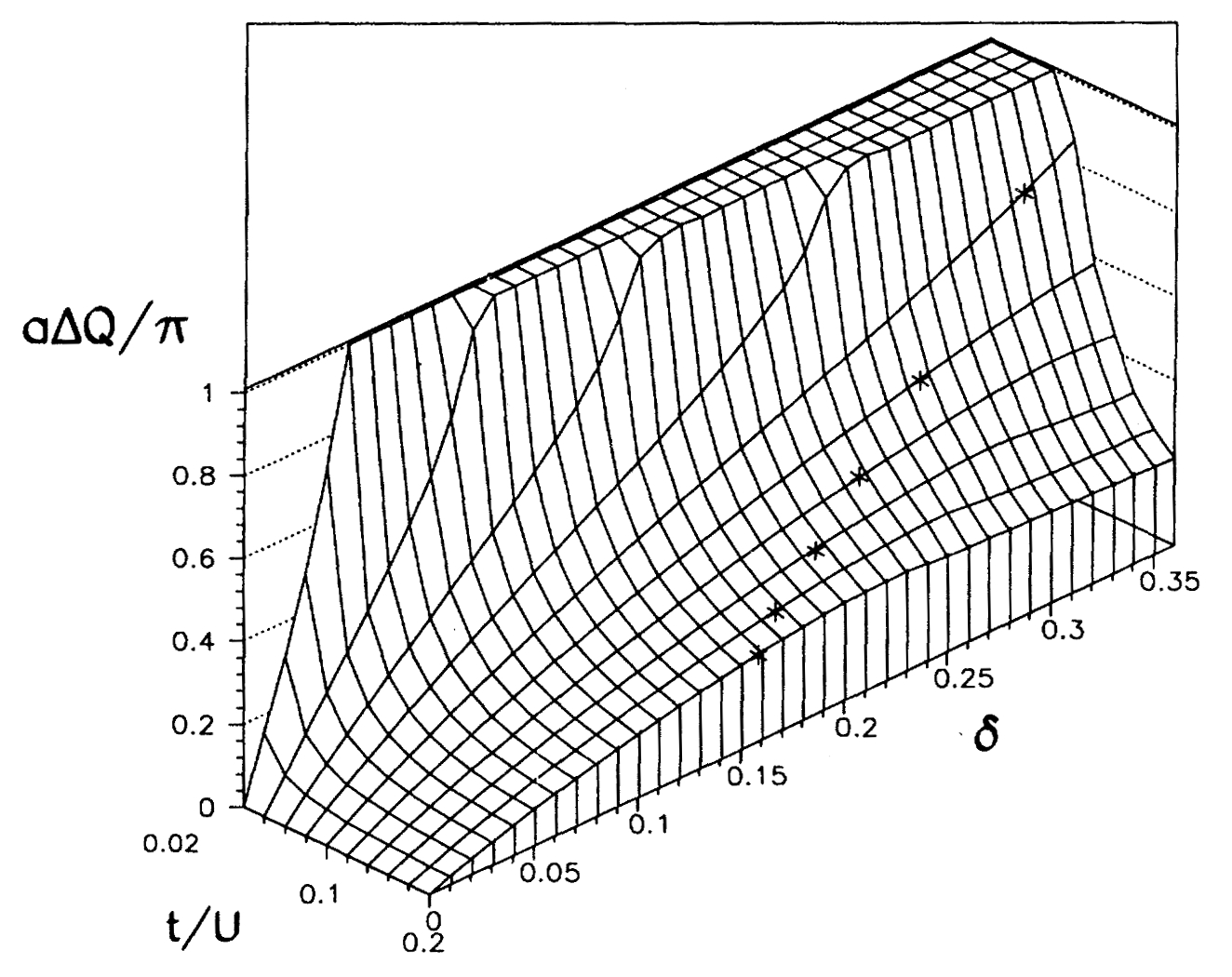}
\includegraphics[width=0.48\textwidth]{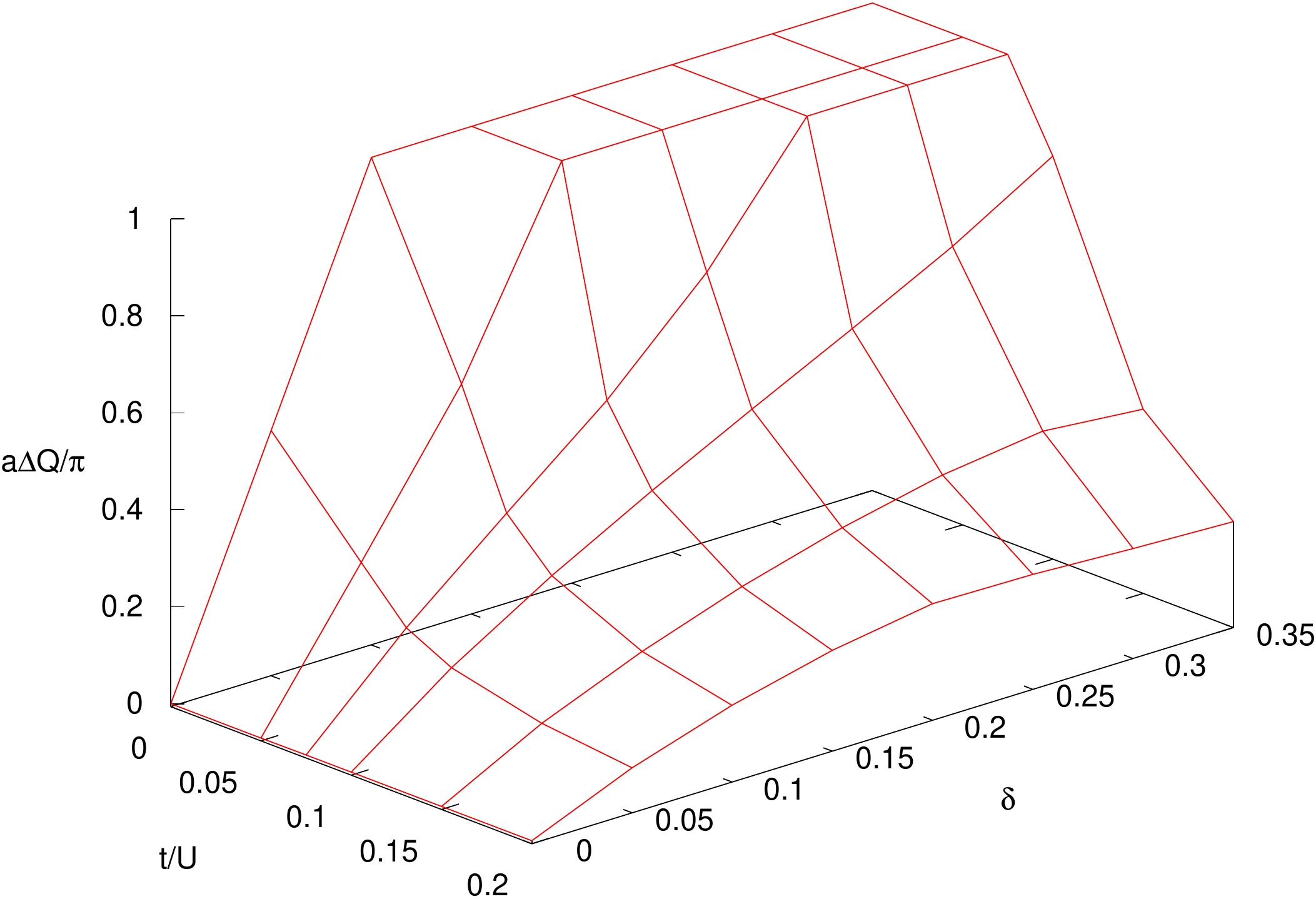} \caption{\label{fig:2}A benchmark of \textit{$\Theta\Phi$} ISS phase diagram
against the results of Ref.\protect\onlinecite{6Arrigoni1991}: (left
panel) Ref.\protect\onlinecite{6Arrigoni1991}, (right panel) \textit{$\Theta\Phi$
}calculations. The axes represent inverse interaction $t/U$, doping
$\delta$ and pitch vector offset from $(\pi,\pi)$ normalized to
$\pi$.}
\end{figure*}

In order to demonstrate the computational efficiency of $\Theta\Phi$, we report in
   Figure~\ref{fig:bench} the nearly perfect scaling of our program's computational time as a function
   of the number of $k$-points in the Brillouin zone. The tests were performed for the single orbital
   Hubbard model on a 2D square lattice. The slope in logarithmic scale implies a quadratic
   dependence in the range of $N_k$ shown,
   as it should be for the integration in 2D.

\subsection{Spiral phases of $\gamma$-Fe}

In the raw of 3\textit{d} materials, Fe falls close to the node separating
the antiferromagnetic metals Cr and Mn with a nearly half-filled \textit{d}
band from strong ferromagnets Co and Ni with a nearly full band. It
is clear, therefore, that the magnetic ordering in Fe should be extremely
sensible to the interatomic distances, bond angles, volume packing
type and other crystallographic details. In fact, it was found experimentally\citep{20Tsunoda1989},
that the magnetic ordering of $\gamma$-\textit{A} phase of Fe precipitates
in Cu is a spin-spiral state propagating with wave vector $Q_{\exp}=2\pi/a(0.1,0,1)$
(Cartesian coordinates).

On the other hand, theoretically, a number of attempts has been made
to tackle this problem (see Refs.\onlinecite{13Deringer2011, 14Maintz2016, 15Tchougreeff2013},
and the references therein). The main theoretical conclusion was that,
in general, there are two minima: one at the $\Gamma X$ \ and the
other at the $\mathit{XW}$ \ lines, while the relative depth of
these minima depends on the unit cell volume: at higher volumes, the
$\Gamma X$ \ line minimum is lower, while at lower volumes, the
$\mathit{XW}$ \ line minimum is more stable. In particular, in the
Ref.\onlinecite{23Marsman2002}, the two minima have the following
positions: $Q_{1}=2\pi/a(0,0,0.6)$ \ and $Q_{2}=2\pi/a(0.2,0,1)$
(Cartesian coordinates). In the internal coordinates, these points
translate into: $Q_{1}=\frac{6}{10}\Gamma X$, while $Q_{2}=\frac{2}{5}\mathit{XW}$.
The rational numbers in these definitions reflect the fact that the
\textbf{\textit{Q}}-vector has to belong to the same finite Brillouin
zone grid as the one used in the DFT calculations ($13\times13\times13$
Monkhorst-Pack in the Ref.\onlinecite{23Marsman2002}). Denser grids
are expected to refine these numbers to some extent.

In this section, we show how this theoretical picture can be reproduced
by using \textit{$\Theta\Phi$}. We derive the hopping parameters
from VASP DFT calculations\citep{24VASP} using the maximally-localized
Wannier orbitals, as implemented in \textsc{wannier90} package\citep{12Mostofi2008}.
We consider $\gamma$-Fe with fcc lattice at several lattice constants:
$a=3.678$\AA, $a=3.583$\AA, $a=3.577$\AA, $a=3.545$\AA, $a=3.510$\AA,
and $a=3.493$\AA, which correspond to decreasing volumes of $V=12.439$\AA\textsuperscript{3}
$V=11.500$\AA\textsuperscript{3}, $V=11.442$\AA\textsuperscript{3},
V=11.138\AA\textsuperscript{3}, $V=10.811$\AA\textsuperscript{3}
and $V=10.655$\AA\textsuperscript{3} respectively. In extracting
the hopping parameters, we utilize VASP paramagnetic DFT calculations
with PBE\citep{25Perdew1996} functional at a $8\times8\times8$ \textit{k}-point
grid, with $E=500\mathit{eV}$ plane-wave cut-off. The quality of
the wannierization procedure employed to obtain the hopping parameters
is similar to the one reported in our previous work (Ref.\onlinecite{11Plekhanov2020},
Appendix C).

In the Refs.\onlinecite{21Knopfle2000,22Bylander1998,23Marsman2002},
the magnetic ordering originated within the LSDA approximation as
a consequence of the Stoner mechanism, where the exchange correlation
functional acts as a ``driving force'', gaining energy from magnetic
polarization. Therefore, in this case, the exchange correlation functional
determines the scale of the total energy landscape as a function of
$\mathbf{Q}$-vector. In our calculations, we use another ``driving
force'', namely the local Heisenberg exchange interaction $J$ which
transforms under spin quantization axis rotation as described in Ref.\onlinecite{11Plekhanov2020}.
This Heisenberg interaction is treated within the mean-field approach.
It is clear that the scale of the relative energy gains will be of
different origin in our case and cannot be directly compared to the
above mentioned references. Nevertheless, we will show that the relative
stability of the two local minima as a function of volume is correctly
reproduced by \textit{$\Theta\Phi$}. We have used $J=-1\mathit{eV}$\ (ferromagnetic
exchange) for Fe. This value was already successfully tested in our
previous \textit{$\Theta\Phi$} calculations of Fe in the Ref.\onlinecite{11Plekhanov2020}.
In addition, this value is constrained from below by the fact that
a value as small as $J=-0.8\mathit{eV}$ \ does not stabilize at
all any magnetic solution, and from above by the observation that
a value as big as $J=-1.2\mathit{eV}$ \ stabilizes a ferromagnetic
solution, which is always more stable than any ISS state. Finally,
we used a full $19\times19\times19$ Monkhorst-Pack $k$-point grid
in our \textit{$\Theta\Phi$} calculations. It was pointed out in
Ref.\onlinecite{23Marsman2002}, that influence of the spin-orbit
coupling on the magnetization in $\gamma$-Fe is very small, that
is why we decided to neglect it in the present work.

Taking into account the above considerations, the Hamiltonian used
in the present section assumes the form:

\begin{align}
H_{\mathrm{Fe}}= & \sum_{i,j,\tau,\alpha,\beta,R}c_{R,i,\alpha}^{\dagger}t_{\mathit{ij}}^{\mathit{\alpha\beta}}(\tau)c_{R+\tau,j,\beta}^{\phantom{\dagger}}\\
+ & \sum_{i,j,\tau,\alpha,\beta,R}J_{i,j}^{\alpha,\beta}(\tau)S_{i,R}^{\alpha}S_{j,R+\tau}^{\beta},
\end{align}
where the hopping matrix $t_{\mathit{ij}}^{\mathit{\alpha\beta}}(\tau)$
is obtained from the wannierization procedure and is transformed under
spin quantization axis rotations as described in the Appendix, $S_{i,R}^{\alpha}$
is the $\alpha$-component of the spin operator on orbital \textit{$i$}
in the cell $R$, while the Heisenberg tensor $J_{i,j}^{\alpha,\beta}(\tau)=\delta(\tau)\delta_{\alpha,\beta}\times J$,
with $J=-1\mathit{eV}$, and the \textit{$i$} and \textit{$j$ }indices
are restricted to the $d$-shell orbitals.

On the other hand, the density matrix in this case has the form summarized
in Table\ref{tab:1} (only the relevant diagonal part of the particle-particle
$\tau=0$ density matrix component is show).

\begin{table*}
\squeezetable

\caption{\label{tab:1}Density matrix table for $\gamma$-Fe.}

\begin{tabular}{|c|c|c|c|c|c|c|c|c|c|}
\hline 
 & $s$ & $p_{z}$ & $p_{x}$ & $p_{y}$ & $d_{z2}$ & $d_{xy}$ & $d_{yz}$ & $d_{x^{2}-y^{2}}$ & $d_{xy}$\tabularnewline
\hline 
\hline 
$\uparrow$ & $\ensuremath{\frac{n_{s}+m_{s}}{2}}$ & $\frac{n_{pz}+m_{pz}}{2}$ & $\frac{n_{px}+m_{px}}{2}$ & $\frac{n_{py}+m_{py}}{2}$ & $\frac{n_{dz2}+m_{dz2}}{2}$ & $\frac{n_{dxz}+m_{dxz}}{2}$ & $\frac{n_{dyz}+m_{dyz}}{2}$ & $\frac{n_{dx2}+m_{dx2}}{2}$ & $\frac{n_{dxy}+m_{dxy}}{2}$\tabularnewline
\hline 
 & $\ensuremath{\frac{n_{s}-m_{s}}{2}}$ & $\frac{n_{pz}-m_{pz}}{2}$ & $\frac{n_{px}-m_{px}}{2}$ & $\frac{n_{py}-m_{py}}{2}$ & $\frac{n_{dz2}-m_{dz2}}{2}$ & $\frac{n_{dxz}-m_{dxz}}{2}$ & $\frac{n_{dyz}-m_{dyz}}{2}$ & $\frac{n_{dx2}-m_{dx2}}{2}$ & $\frac{n_{dxy}-m_{dxy}}{2}$\tabularnewline
\hline 
\end{tabular}
\end{table*}

Here, for the sake of brevity, $18$ diagonal matrix elements are
arranged in a $2\times9$ table with the orbital indices running along
the columns and the spin components running along the rows. The order
of the orbitals corresponds to the one established in \textsc{wannier90}.
Finally, the matrix elements are expressed in terms of the parameters
$n_{i}$ and $m_{i}$, so that, for example, $n_{s}$ and $m_{s}$
\ are the occupation and the magnetization of the \textit{s }orbital
respectively.

\textit{$\Theta\Phi$} results for $\gamma$-Fe are shown in Figure
3. The total energy dependence in our calculations is, indeed, very
sensible to the unit cell volume. Starting from $a=3.493$ \AA, the
relative energy gain of the $\mathbf{Q}_{1}$ spin spiral gradually
grows up to $a=3.545$ \AA, and then, as the volume increases further,
starts to decrease until, at $a=3.678$ \AA, a situation realizes
when the ferromagnetic state at $\mathbf{Q}=0$ becomes the most stable.
The experimental volume $V=11.442$ \AA\textsuperscript{3}, corresponding
to $a=3.577$ \AA, is close to the maximum energy gain tested in
the present paper: $0.1\mathit{eV}$. The $\mathbf{Q}_{2}$ spin spiral
is not reproduced in our calculations in the sense that the broad
minimum of the energy in the $\mathit{XW}$ interval is realized at
the $W$ point or in its vicinity, although the energy difference
between the $\mathbf{Q}_{1}$ spin spiral and the whole $\mathit{XW}$
line at the experimental volume is extremely small.

\begin{figure*}
\includegraphics[width=0.48\textwidth]{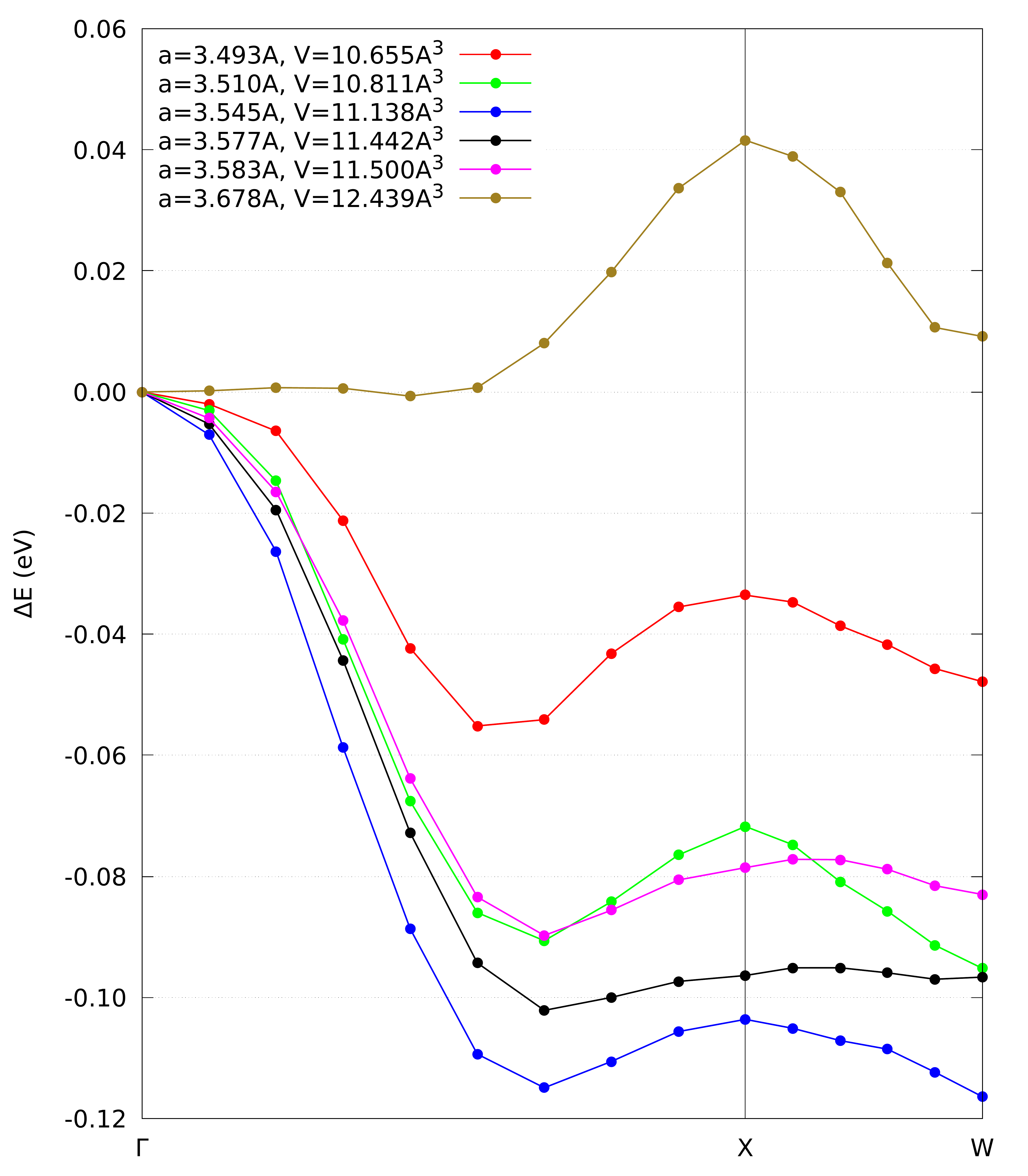} \includegraphics[width=0.48\textwidth]{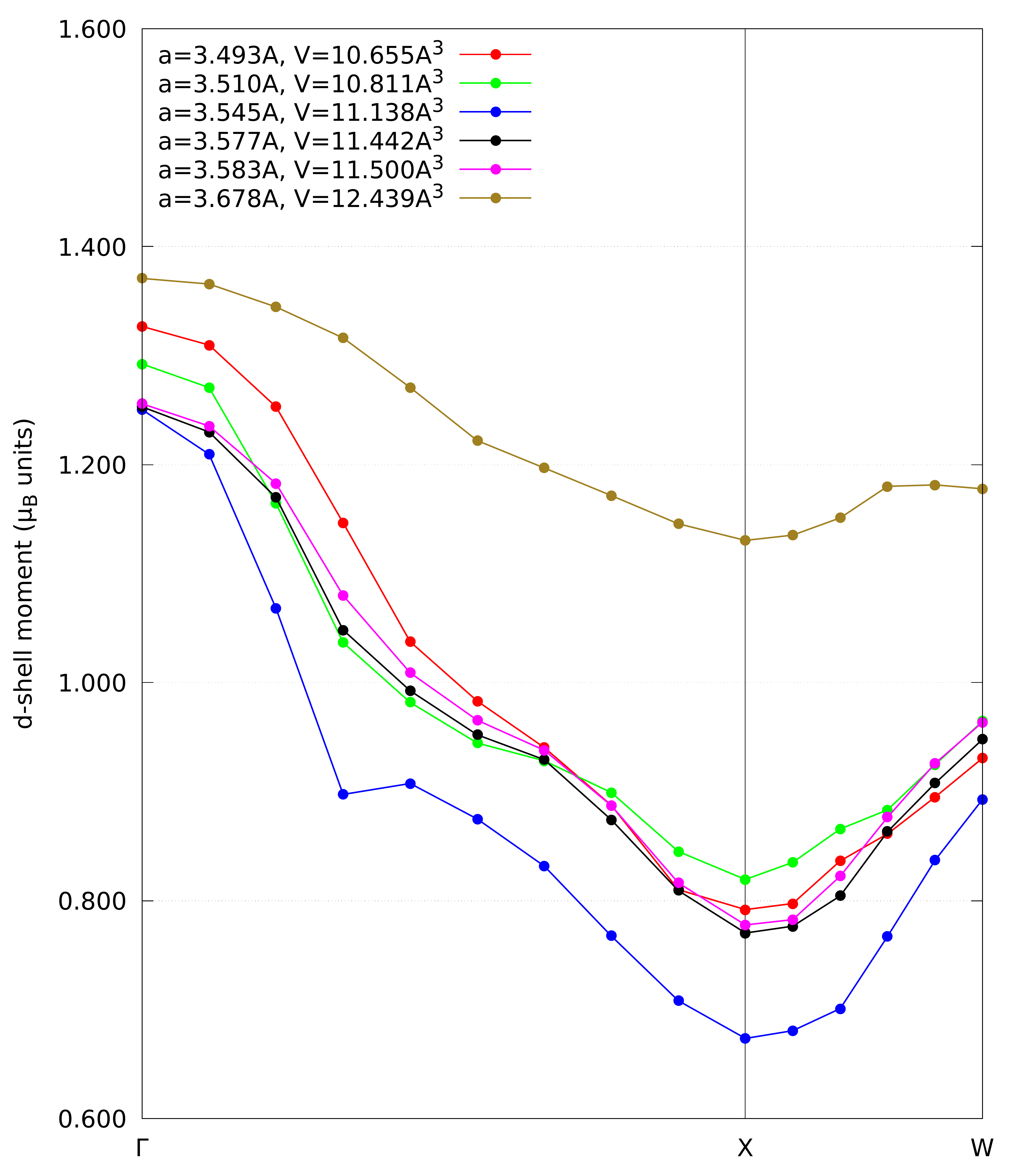}
\caption{\label{fig:3}Left panel: A scan on pitch vector along the $\Gamma\to X\to W$
path of the spin spiral state energy gain with respect to the ferromagnetic
state for different unit cell volumes to compare against the results
of Ref.\protect\onlinecite{23Marsman2002}. Right panel: The total
\textit{d}-shell magnetization along the same path for different volumes.
The black lines correspond to the experimental volume.}
\end{figure*}

In our calculations, the \textit{d}-shell magnetic moment of the spiral
phase decreases as a function of $\mathbf{Q}$ at $\Gamma X$ line
and increases at $\mathit{XW}$ line, having a minimum at $X$, as
shown in Figure\ref{fig:3}.The absolute value of the moment, approximately
one $\mu_{B}$, is similar to the values reported in the Ref.\onlinecite{23Marsman2002}.
The energy gain of the $\mathbf{Q}_{1}$ spin spiral phase with respect
to the $\mathbf{Q}=0$ phase is approximately twice larger than the
one reported in the Refs.\onlinecite{21Knopfle2000,23Marsman2002},
and this difference can be ascribed to a completely different mechanism
which stabilizes the magnetic phase, as mentioned above.

\section[Discussion and Conclusions]{Discussion and Conclusions}

In the present paper, we have shown how \textit{$\Theta\Phi$} program
can be easily used to calculate the properties of incommensurate spin
density waves. We have performed the benchmark of our results against
those of Ref.\onlinecite{6Arrigoni1991} and Ref.\onlinecite{23Marsman2002}
and shown the excellent agreement. In particular, for the single-orbital
Hubbard model we have shown that the ISS state can be stabilized in
a wide range of the on-site repulsion $U$, and the pitch vector off-set
from the commensurability varies from zero to one as a function of
$U$ and doping $\delta$. In addition, the \textit{$\Theta\Phi$}
minimization procedure stably locates the total energy minimum as
a function of the pitch vector $\mathbf{Q}$.

In the case of the ISS in $\gamma$-Fe, we have shown that it can
be successfully stabilized and has the energy lower than the ferromagnetic
state, at least along the path $\Gamma\to X\to W$. Although in our
calculations the interaction term, was completely different from the
Refs.\onlinecite{21Knopfle2000,23Marsman2002} our results are rather
similar to the ones reported in there. Indeed, in DFT the interaction
comes from the exchange functional, which, in turn, is obtained by
fitting the homogeneous electron gas Monte Carlo simulations, while
in our calculations, the full ($x,y,z$-components) Heisenberg interaction
on \textit{d}-orbitals, supplied with the spin quantization axis rotation
formulas was used. The only parameter in our calculations -- The
Heisenberg exchange $J$-- is close to the values routinely used
for iron and is constrained from below and above by the absence of
magnetic solution and the instability with respect to the ferromagnetic
phase respectively. Taking into account these differences, our results
are in surprisingly good agreement with the previously published ones:
the sensitivity to the unit cell volume, the stability of the $\mathbf{Q}_{1}$
state, the order of magnitude of the energy gain, the value of the
magnetic moment. The only feature not reproduced in our calculations
is the absolute stability of the $\mathbf{Q}_{2}$ state, although
the energy difference between the $\mathbf{Q}_{1}$ and the $\mathbf{Q}_{2}$
states is very small. This discrepancy is currently under investigation
and will be explained in a later work.

Additionally, in this paper, we have proposed a general approach for
the spin quantization axis rotation of the most problematic Hamiltonian
part -- kinetic energy. This approach only requires the hopping matrix
to be Hermitian, without demanding that the off-diagonal spin-flip
terms like spin-orbit coupling be small and treated as perturbations.
Such an approach will allow us to efficiently calculate the ISS in
heavy elements with sizable spin-orbit coupling like lanthanides and
actinides.

Within \textit{$\Theta\Phi$}, both single-orbital case and a general
multi-orbital one can be routinely treated as shown here and in Refs.\onlinecite{10Plekhanov2017,11Plekhanov2020}.
This paves the way to the computationally cheap calculations of ISS
in complex multi-orbital systems with the `\textit{ab-initio}' predictive
power. These calculations can be further combined with the large-scale
material search codes (see \textit{e.g.} Ref.\onlinecite{26Oganov2006})
in order to perform the large scale material search with the given
functional properties.




\appendix

\section{Rotating the local quantization axes}

In this Appendix, we briefly review the method of local quantization
axes rotation, originally formulated in Refs.\onlinecite{6Arrigoni1991,8Sandratskii1991,9Sandratskii1991,11Plekhanov2020,27Barreteau2016}.

The most general form of spin-dependent hopping between an orbital
$i$ and an orbital $j$ is:

\begin{equation}
t_{\mathit{ij}}(\tau)=\left(\begin{matrix}t_{\mathit{ij}}^{\uparrow\uparrow}(\tau) & t_{\mathit{ij}}^{\uparrow\downarrow}(\tau)\\
t_{\mathit{ij}}^{\downarrow\uparrow}(\tau) & t_{\mathit{ij}}^{\downarrow\downarrow}(\tau)
\end{matrix}\right).
\end{equation}

It obeys the hermiticity condition:

\[
t_{\mathit{ij}}(\tau)=t_{\mathit{ji}}^{\dagger}(-\tau).
\]
On the other hand, the rotation operator (around $y$-axis) reads
as:

\[
\Omega_{\tau}=e^{-i\frac{(\mathbf{Q},\tau)}{2}\sigma_{y}}.
\]

Here, we rotate the spin-quantization axes around y-axis, that is
why $\sigma_{y}$ \ appears in $\Omega_{\tau}$. The other choices
would be rotations around \textit{x}- and \textit{z}-axis and could
be simply accounted for by substituting $y\rightarrow x$ \ or $y\rightarrow z$.
For the moment, we leave apart the individual orbital rotations $\Omega_{i}$\ (which
are also implemented in \textit{$\Theta\Phi$}) since they do not
interfere with the lattice translations, and we will take them into
account in the final answer. The rotated-by-the-super-structure-vector-$\mathbf{Q}$
kinetic energy reads as:
\begin{widetext}
\begin{equation}
\widetilde{T}=\sum_{i,j,\tau,\alpha,\beta,R}c_{R,i,\alpha}^{\dagger}\left(e^{i\frac{(\mathbf{Q},R)}{2}\sigma_{y}}t_{\mathit{ij}}(\tau)e^{-i\frac{(\mathbf{Q},(R+\tau))}{2}\sigma_{y}}\right)_{\alpha,\beta}c_{R+\tau,j,\beta}^{\phantom{\dagger}}.
\end{equation}
\end{widetext}

We convert the summation on $R$ into the summation on $k$ and $k^{\prime}$:

\begin{align}
\widetilde{T} & =\frac{1}{N}\sum_{i,j,\tau,\alpha,\beta,R,k,k^{\prime}}e^{-ik^{\prime}\tau}e^{i(k-k^{\prime})R}c_{k,i,\alpha}^{\dagger}\\
 & \times\left(e^{i\frac{(\mathbf{Q},R)}{2}\sigma_{y}}t_{\mathit{ij}}(\tau)e^{-i\frac{(\mathbf{Q},(R+\tau))}{2}\sigma_{y}}\right)_{\alpha,\beta}c_{k^{\prime},j,\beta}^{\phantom{\dagger}}
\end{align}

If we define the matrix $L_{\mathit{ij}}(k-k^{\prime},\tau)$:

\begin{equation}
L_{\mathit{ij}}(k-k^{\prime},\tau)=\frac{1}{N}\sum_{R}e^{i(k-k^{\prime})R}e^{i\frac{(\mathbf{Q},R)}{2}\sigma_{y}}t_{\mathit{ij}}(\tau)e^{-i\frac{(\mathbf{Q},R)}{2}\sigma_{y}},
\end{equation}

then the kinetic energy expression simplifies as follows:

\begin{align}
\widetilde{T}=\sum_{i,j,\tau,\alpha,\beta,k,k^{\prime}}e^{-ik^{\prime}\tau}c_{k,i,\alpha}^{\dagger}L_{\mathit{ij}}(k-k^{\prime},\tau)e^{-i\frac{(\mathbf{Q},\tau)}{2}\sigma_{y}}c_{k^{\prime},j,\beta}\label{eq:Ttilde}\\
=\sum_{i,j,\tau,\alpha,\beta,k,k^{\prime}}e^{-\mathit{ik\tau}}c_{k,i,\alpha}^{\dagger}e^{-i\frac{(\mathbf{Q},\tau)}{2}\sigma_{y}}L_{\mathit{ij}}(k-k^{\prime},\tau)c_{k^{\prime},j,\beta}.\nonumber 
\end{align}

Let us consider the matrix $L_{\mathit{ij}}(k-k^{\prime},\tau)$.
First of all, if the matrix $t_{\mathit{ij}}(\tau)$ \ is proportional
to identity, or to $\sigma_{y}$ then the rotations on the left and
right hand side cancel each other and $L_{\mathit{ij}}(k-k^{\prime},\tau)=\delta(k-k^{\prime})t_{\mathit{ij}}(\tau)$
and we obtain the result outlined in the Ref.\onlinecite{11Plekhanov2020}.
Although, there is still a room for the non-trivial results, the most
physically interesting situation is when $t_{\mathit{ij}}(\tau)$
does not commute with $\sigma_{y}$, spin-orbit coupling being an
important example. In the general case, we can write:

\[
\Omega_{\tau}=\cos\frac{(\mathbf{Q},\tau)}{2}-i\sigma_{y}\sin\frac{(\mathbf{Q},\tau)}{2},
\]
so that after some simplification we have:
\begin{widetext}
\begin{equation}
L_{\mathit{ij}}(k-k^{\prime},\tau)=\frac{1}{2N}\sum_{R}e^{i(k-k^{\prime})R}\times\left\{ t_{\mathit{ij}}(\tau)+\sigma_{y}t_{\mathit{ij}}(\tau)\sigma_{y}+\left(t_{\mathit{ij}}(\tau)-\sigma_{y}t_{\mathit{ij}}(\tau)\sigma_{y}\right)\cos(\mathbf{Q},R)+i\left[\sigma_{y},t_{\mathit{ij}}(\tau)\right]\sin(\mathbf{Q},R)\right\} .
\end{equation}
\end{widetext}

At this point, we can do the Fourier transform in the above expression,
but the problem now is that the resulting expression will not be diagonal
in $k$-space, \textit{i.e.} there will be terms with $k$ and $k\pm\mathbf{Q}$
connected!

Till now, the formalism is completely general and other cases of rotation
around \textit{e.g.} $x$ or $y$ axes can be performed by substituting
$y\rightarrow x$, or $y\rightarrow z$. The case of rotation around
$z$ \ axis is particularly instructive since $\sigma_{z}$ is diagonal.
In the most general case we can present:

\begin{align}
t_{\mathit{ij}}(\tau) & =\left(\begin{matrix}a_{\mathit{ij}}(\tau) & c_{\mathit{ij}}(\tau)\\
d_{\mathit{ij}}(\tau) & b_{\mathit{ij}}(\tau)
\end{matrix}\right)=t_{\mathit{ij}}^{\parallel}+t_{\mathit{ij}}^{\perp}\\
 & =\left(\begin{matrix}a_{\mathit{ij}}(\tau) & 0\\
0 & b_{\mathit{ij}}(\tau)
\end{matrix}\right)+\left(\begin{matrix}0 & c_{\mathit{ij}}(\tau)\\
d_{\mathit{ij}}(\tau) & 0
\end{matrix}\right).
\end{align}

The first matrix will commute with $\sigma_{z}$ and will only contribute
to the first term (independent on $\mathbf{Q}$) in $L_{\mathit{ij}}(k-k^{\prime},\tau)$.
For $t_{\mathit{ij}}^{\perp}$ we can explicitly work out:

\begin{align}
\sigma_{z}t_{\mathit{ij}}^{\perp}(\tau)\sigma_{z} & =-t_{\mathit{ij}}^{\perp}(\tau)\\
\left[\sigma_{z},t_{\mathit{ij}}^{\perp}(\tau)\right] & =2\left(\begin{matrix}0 & c_{\mathit{ij}}(\tau)\\
-d_{\mathit{ij}}(\tau) & 0
\end{matrix}\right).
\end{align}

Therefore, $L_{\mathit{ij}}(k-k^{\prime},\tau)$ will become:

\begin{align}
L_{\mathit{ij}}(k-k^{\prime},\tau) & =t_{\mathit{ij}}^{\parallel}(\tau)\delta(k-k^{\prime})\\
 & +t_{\mathit{ij}}^{\perp}(\tau)\left(\begin{matrix}\delta(k-k^{\prime}-\mathbf{Q}) & 0\\
0 & \delta(k-k^{\prime}+\mathbf{Q})
\end{matrix}\right).\nonumber 
\end{align}

At this point we can introduce the Fourier transforms:

\[
\sum_{\tau}t_{\mathit{ij}}(\tau)e^{-\mathit{ik\tau}}=\left(\begin{matrix}a_{\mathit{ij}}(k) & c_{\mathit{ij}}(k)\\
c_{\mathit{ij}}^{\star}(k) & b_{\mathit{ij}}(k)
\end{matrix}\right),
\]
where we have used the hermiticity condition:

\[
\sum_{\tau}d_{\mathit{ij}}(\tau)e^{-\mathit{ik\tau}}=\sum_{\tau}c_{\mathit{ij}}^{\star}(-\tau)e^{-\mathit{ik\tau}}.
\]
Substituting $L_{\mathit{ij}}(k-k^{\prime},\tau)$ into Eq.\eqref{eq:Ttilde}
and changing $\sigma_{y}\rightarrow\sigma_{z}$, we obtain, after
simplifications:

\begin{align}
 & \widetilde{T}_{z}(\mathbf{Q})=\sum_{i,j,k}\left(c_{k,i,\uparrow}^{\dagger},c_{k+\mathbf{Q},j,\downarrow}^{\dagger}\right)\times\\
 & \left(\begin{matrix}a_{\mathit{ij}}(k+\frac{\mathbf{Q}}{2}) & c_{\mathit{ij}}(k+\frac{\mathbf{Q}}{2})\\
c_{\mathit{ij}}^{\star}(k+\frac{\mathbf{Q}}{2}) & b_{\mathit{ij}}(k+\frac{\mathbf{Q}}{2})
\end{matrix}\right)\left(\begin{matrix}c_{k,i,\uparrow}^{\phantom{\dagger}}\\
c_{k+\mathbf{Q},j,\downarrow}^{\phantom{\dagger}}
\end{matrix}\right).
\end{align}
The rotation around $y$-axis can be evaluated similarly, although
the calculations in that case are more cumbersome. We report below
the final answer:

\begin{align}
 & \widetilde{T}_{y}(\mathbf{Q})=\sum_{i,j,k}\left(c_{k,i,\uparrow}^{\dagger},c_{k+\mathbf{Q},j,\downarrow}^{\dagger}\right)\times\\
 & \left(\begin{matrix}\alpha_{\mathit{ij}}(k+\frac{\mathbf{Q}}{2}) & \gamma_{\mathit{ij}}(k+\frac{\mathbf{Q}}{2})\\
\gamma_{\mathit{ij}}^{\star}(k+\frac{\mathbf{Q}}{2}) & \beta_{\mathit{ij}}(k+\frac{\mathbf{Q}}{2})
\end{matrix}\right)\left(\begin{matrix}c_{k,i,\uparrow}^{\phantom{\dagger}}\\
c_{k+\mathbf{Q},j,\downarrow}^{\phantom{\dagger}}
\end{matrix}\right).
\end{align}
where:

\begin{align}
\alpha_{\mathit{ij}}(k) & =\phantom{-}\frac{a_{\mathit{ij}}(k)+b_{\mathit{ij}}(k)}{2}+\mathrm{Im\,}c_{\mathit{ij}}(k)\\
\beta_{\mathit{ij}}(k) & =\phantom{-}\frac{a_{\mathit{ij}}(k)+b_{\mathit{ij}}(k)}{2}-\mathrm{Im\,}c_{\mathit{ij}}(k)\\
\gamma_{\mathit{ij}}(k) & =-\frac{a_{\mathit{ij}}(k)+b_{\mathit{ij}}(k)}{2}-i\mathrm{Re\,}c_{\mathit{ij}}(k)
\end{align}

It is easy to see that $\widetilde{T}_{z}(\mathbf{Q})$ \ and $\widetilde{T}_{y}(\mathbf{Q})$
are different representations of the same operator and, indeed, are
related by a unitary transformation $S$:

\[
S=\frac{1}{\sqrt{2}}\left(\begin{matrix}i & -i\\
1 & 1
\end{matrix}\right),
\]
so that:

\[
S^{\dagger}\left(\begin{matrix}a_{\mathit{ij}}(k) & c_{\mathit{ij}}(k)\\
c_{\mathit{ij}}^{\star}(k) & b_{\mathit{ij}}(k)
\end{matrix}\right)S=\left(\begin{matrix}\alpha_{\mathit{ij}}(k) & \gamma_{\mathit{ij}}(k)\\
\gamma_{\mathit{ij}}^{\star}(k) & \beta_{\mathit{ij}}(k)
\end{matrix}\right).
\]

It can be seen, that the two ways to rotate the quantization axis
are fully equivalent. Since rotation around \textit{z}-axis features
somehow simpler formulas, we will carry out further calculations assuming
\textit{z}-axis rotation.

Putting back the individual orbital rotations $\Omega_{i}$, we can
assemble the final rotation formula for the multi-orbital case as
follows:
\begin{widetext}
\begin{equation}
\widetilde{T}{}_{z}(\mathbf{Q})=\sum_{i,j,k}\left(c_{k,i,\uparrow}^{\dagger},c_{k+\mathbf{Q},j,\downarrow}^{\dagger}\right)\Omega_{i}^{\dagger}\left(\begin{matrix}a_{\mathit{ij}}(k+\frac{\mathbf{Q}}{2}) & c_{\mathit{ij}}(k+\frac{\mathbf{Q}}{2})\\
c_{\mathit{ij}}^{\star}(k+\frac{\mathbf{Q}}{2}) & b_{\mathit{ij}}(k+\frac{\mathbf{Q}}{2})
\end{matrix}\right)\Omega_{j}\left(\begin{matrix}c_{k,i,\uparrow}^{\phantom{\dagger}}\\
c_{k+\mathbf{Q},j,\downarrow}^{\phantom{\dagger}}
\end{matrix}\right).
\end{equation}
\end{widetext}

In this formula, the states with different $k$-vectors are mixed.
This breaks the translational invariance of the crystal and makes
it impossible to effectively treat ISS states. The way out of this
situation is to shift the down-spin electronic states by the vector
$\mathbf{Q}$ so that in the new notation the Fockian matrix will
be diagonal in $k$-space:

\begin{align}
 & \widetilde{T}_{z}(\mathbf{Q})=\sum_{i,j,k}\left(c_{k,i,\uparrow}^{\dagger},\widetilde{c}_{k,j,\downarrow}^{\dagger}\right)\Omega_{i}^{\dagger}\times\\
 & \left(\begin{matrix}a_{\mathit{ij}}(k+\frac{\mathbf{Q}}{2}) & c_{\mathit{ij}}(k+\frac{\mathbf{Q}}{2})\\
c_{\mathit{ij}}^{\star}(k+\frac{\mathbf{Q}}{2}) & b_{\mathit{ij}}(k+\frac{\mathbf{Q}}{2})
\end{matrix}\right)\Omega_{j}\left(\begin{matrix}c_{k,i,\uparrow}^{\phantom{\dagger}}\\
\widetilde{c}_{k,j,\downarrow}^{\phantom{\dagger}}
\end{matrix}\right),
\end{align}
where $\widetilde{c}_{k,j,\downarrow}^{\phantom{\dagger}}=c_{k+\mathbf{Q},j,\downarrow}^{\phantom{\dagger}}$.

For what regards the interaction term, we consider here only the generalized
Heisenberg term, as defined in the Ref.\onlinecite{11Plekhanov2020}:

\[
H_{J}=\sum_{i,j,R,\tau,\alpha,\beta}J_{i,j}^{\alpha,\beta}(\tau)S_{i,R}^{\alpha}S_{j,R+\tau}^{\beta},
\]
where $S_{i,R}^{\alpha}=\frac{1}{2}\sum_{s,s^{\prime}}c_{R,i,s}^{\dagger}\sigma_{s,s^{\prime}}^{\alpha}c_{R,i,s^{\prime}}^{\dagger}$
is the operator of the $\alpha$-s component of the spin of the orbital
$i$ in the cell \textit{$R$}. The transformation of the exchange
coupling $J_{i,j}^{\alpha,\beta}(\tau)$ under the spin quantization
axis rotation is reported in the Ref.\onlinecite{11Plekhanov2020}.

\bibliographystyle{apsrev4-1}

%

\end{document}